\documentclass{aa}             

\usepackage{graphicx}

\begin{document}

\title{BL Lac contribution to the extragalactic gamma-ray background}

 \author{T. M. Kneiske\inst{1,2} \and K. Mannheim\inst{1}}

 \institute{Universit\"at W\"urzburg, Am Hubland, 97057 W\"urzburg, Germany
 \and Department of Physics and Mathematical Physics, University of Adelaide, SA 5005, Australia}
 \offprints{T.M. Kneiske \email{tkneiske@physics.adelaide.edu.au}}

 \date{Received/Accepted}

 \abstract{Very high energy gamma-rays from blazars
traversing cosmological distances through the metagalactic radiation
field can convert into electron-positron pairs in photon-photon collisions.
The converted gamma-rays initiate electromagnetic cascades
driven by inverse-Compton scattering off the microwave background photons. 
Using a model for the time-dependent metagalactic radiation  field consistent with all
currently available far-infrared-to-optical data, we calculate
the cascade contribution from faint, unresolved high- and low-peaked blazars to the extragalactic
gamma-ray background as measured by EGRET.
For low-peaked blazars, we adopt a spectral index consistent with the mean
spectral index of EGRET detected blazars, and the EGRET luminosity function.
For high-peaked blazars,
 we adopt template spectra matching prototype sources observed
with air-Cherenkov telescopes up to 30 TeV, and a luminosity function
based on X-ray measurements.
The low number of $\sim 20$ for nearby high-peaked blazars
with a flux exceeding $10^{-11}~\rm cm^{-2}s^{-1}$ above 300~GeV inferred from the luminosity function
is consistent with the results from air-Cherenkov telescope observations.
Including the cascade emission from higher redshifts,
the total high-peaked blazar contribution to the observed gamma-ray background at GeV energies
can account for $\sim30\%$.}
\maketitle

\section{Introduction}
Isotropic, diffuse
background radiation presumably due to faint, unresolved extragalactic sources
has been observed in nearly all energy bands.
The confirmation of an extragalactic gamma-ray background by EGRET (Energetic
Gamma-Ray Experiment Telescope) on board the Compton Gamma Ray Observatory
has extended the spectrum up to
an energy of $\sim$50~GeV.
A first analysis of the data resulted in
a total flux of $(1.45\pm 0.05) \cdot 10^{-5}$ photons cm$^{-2}$ s$^{-1}$
sr$^{-1}$ above 100~MeV and a spectrum that could be fitted by
a power law with an spectral index of $-2.10\pm 0.03$ (Sreekumar et al 1998).
These values are strongly dependent on the
foreground emission model which is subtracted from the observed
intensity to obtain the extragalactic residual.
Since, using the foreground model by Hunter et al (1997), a residual GeV halo remained after subtraction (in
addition to the isotropic extragalactic background), the foreground model had to be
improved. This led to a new analysis of the EGRET data, and a new result for the
extragalactic background spectrum, now showing a dip at GeV energies
and an overall weaker intensity of
$(1.14\pm 0.12) \cdot 10^{-5}$ photons cm$^{-2}$ s$^{-1}$ sr$^{-1}$
(Strong, Moskalenko \& Reimer 2004).
This new result can help us to understand the origin of the extragalactic
background radiation.

Since EGRET detected a large number of extragalactic
gamma-ray sources belonging to the blazar class of AGN, a reasonable assumption is that the gamma background
is produced by unresolved AGN.
Stecker \& Salamon (1996) were able to explain 100\% of the background
by including the effect of variabiliy.
But the resulting number of faint, nearby blazars was too low
compared to the numbers detected by EGRET (Chiang \& Mukherjee 1998).
Using a gamma-ray luminosity function from EGRET data,
Chiang \& Mukherjee (1998) found that only
25\% to 50\% of the gamma-ray  background could be explained by blazars.
This result was questioned by Stecker (2001) who argued that 
a statistically independent analysis is contrary to his assumption 
of a correlated radio and gamma-ray emission and introduces a bias.
Giommi et al. (2006) estimate the gamma-ray background contribution of blazars from
deep blazar counts in the radio band, from multi-frequency surveys, and from interpolations
of data using synchrotron-self-Compton models.
They found that blazars possibly can contribute
100\% in the $\propto 0.5 - 50$~MeV band. Since their calculation
over-predicts the observations at energies ($E>100$~MeV) by a large factor they conclude that
the duty cycle of blazars must be rather low, and this is consistent with the results
from long-term monitoring of TeV blazars such as Mrk421 or Mrk501. Due to the gap in sensitivity
in the 100~keV to 100 MeV band, the spectral shape of blazars between medium energy gamma rays
and high energy gamma rays is poorly constrained by observations.

Here extending the
existing models we assume a population of BL Lacertae objects (BL Lac) with a spectral
energy distribution such that their flux at EGRET energies is generally too low
to be detected, while their very high energy gamma ray flux is strong.
Most of these sources are at redshifts high enough for pair attenuation with
the UV-IR metagalactic radiation field (MRF) to
take place. A significant part of their VHE emission is reprocessed by
inverse Compton cascades (Protheroe 1986, Maciolek-Niedzwiecki,
Zdzarski \& Coppi 1995, Protheroe \& Stanev 1993). For a low extragalactic magnetic
field strength ($<10^{-17}$~G) the emission contributes to the source spectrum
(Dai et al. 2002, Fan, Dai \& Wei 2004). Assuming a much higher magnetic field of $10^{-9}$~G
the electrons and positrons become isotropized. The photons produce
the so-called gamma-ray halo (Aharonian, Coppi \& V\"olk 1994).
The inverse Compton emission of blazars can contribute to the GeV gamma-ray background
(Coppi \& Aharonian 1997).

The paper is organized as follows: in the next section we present the set of equations used
to calculate the gamma-ray background including the component from undetected
sources and, assuming no extragalactic magnetic field, from cascade emission.
In Sections 3 and 4 we give the choice of parameters for two populations of AGN, the
undetected EGRET blazars and high peaked BL Lacs. The results are shown in Section 5, followed
by a discussion in Section 6.

In this paper we use a Hubble constant of $H_0=71$km s$^{-1}$ Mpc$^{-1}$ and
a flat Universe with the cosmological parameters $\Omega_{\mathrm M}$=0.3 and $\Omega_\Lambda$=0.7.

\section{Gamma-ray background}

The emission in units of [GeV$^{-1}$ cm$^{-2}$ s$^{-1}$ sr$^{-1}$] of a population of
unresolved gamma-ray sources to the extragalactic gamma-ray background
can be described by

\begin{eqnarray}
\frac{dN}{dE_\gamma\ d\Omega} & = & \frac{1}{4\pi} \int^{z_m}_0  \frac{dV_c}{dz_s}
\int^{\infty}_{L_{\mathrm m}} \frac{dN}{dV\ dL} N_0 \cdot \nonumber \\
& \cdot & \left[\frac{dN^i}{dE_\gamma}(z_s)+\frac{dN^c}{dE_\gamma}(z_s)\right]\ e^{-\tau_{\gamma \gamma}(z_s)}\ \mathrm{d}L
 \ \mathrm{d}z_s,
 \ \ \ \
 \label{eq:gammaback}
\end{eqnarray}
with $dN^i/dE_\gamma(z_s, L)$ as the intrinsic gamma-ray flux and
the cascade emission $dN^c/dE_\gamma(z_s, L)$
from a source at redshift $z_s$, tacitly assuming that the cascade contributes
at the redshift of the source which is the case for all
pair creation optical depths $\tau_{\gamma\gamma}\gg1$.
This assumption is valid for most of the absorbed spectra, which we assume to
generically extend to 30~TeV, since the majority of sources at redshifts $z\sim1$ have $\tau_{\gamma\gamma}\gg1$
above $50$~GeV.
The cosmological volume element is $dV_c/dz$,
$L_{m}$ is the total luminosity of the weakest source, $dN/(dV\ dL)$
the luminosity function and $N_0$ is the normalization constant for a single source spectrum
defined as a function of the total intrinsic luminosity $L$

\begin{equation}
N_0 = \frac{L}{4\pi d_l^2(z_s)}
\left(\int_{E_{\mathrm{min}}}^{E_{\mathrm{max}}}E_{\gamma}'
\frac{dN^i}{dE_\gamma'}(z_s, L)\ dE_\gamma'\right)^{-1}.
\label{eq:Ltot}
\end{equation}
Here the limits $E_\mathrm{min}$ and $E_\mathrm{max}$ depend on the definition
of the energy range of the
luminosity used in the luminosity function.

\subsection{Extragalactic gamma-ray absorption}
The high energy photons ($>$~20~GeV) of cosmic gamma-ray sources undergo
interactions with the UV-FIR metagalactic radiation field (MRF).
Therefore gamma-ray photons are absorbed due to pair-production (Kneiske et al. 2004
and references therein).

We use a broken power-law for the intrinsic spectrum in the energy range from 100~MeV to
30~TeV
\begin{equation}
\frac{dN^i}{dE_\gamma}(z_s) \propto E_\gamma^{-\alpha}
 \label{eq:Pow}
\end{equation}
 with $\alpha=\alpha_1$ for $E_\gamma'>E_{\mathrm peak}'$
and $\alpha=\alpha_2$ for $E_\gamma'\leq E_{\mathrm peak}'$.

The two slopes and peak energy will be determined by the spectrum of the observed sources.

\begin{figure}

\vspace{0.7cm}

  \includegraphics[height=.24\textheight]{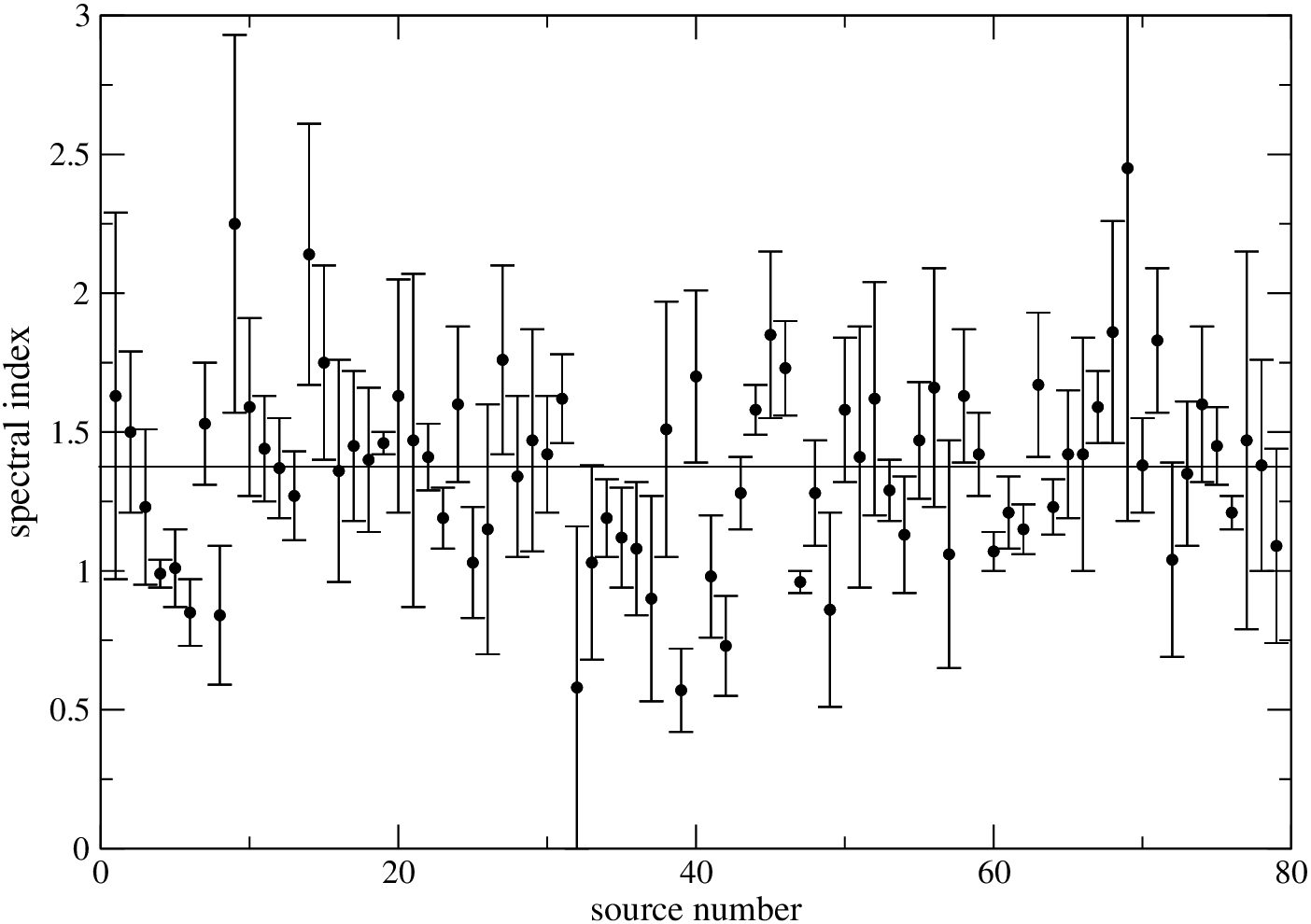}
  \caption{Spectral indices of blazars in the 3rd EGRET catalog (mean index
$1.37\pm0.04$)}
  \label{fig:meanEGRET}
  
\end{figure}

\subsection{Cascade contribution}
The following analytical approximation, similar to the one in
Fan, Dai \& Wei (2004), used to calculate the
particle flux of a pair-cascade following the pair production process
from a gamma-ray source at redshift $z_s$.  The cascade term in
Eq.~(\ref{eq:gammaback}) accounts
for the first generation of electrons produced in photon-photon pair
production. Corrections for the second generation of electrons is negligible
since the contribution is very small for the chosen gamma energy range.
For all spectra we assume a maximum energy of 30~TeV, so the inverse
Compton scattering takes place in the Thomson regime only.
The choice of this energy
limit is motivated by the highest energies observed in BL Lac spectra.
Note that the effect of a higher maximum energy has been
studied in Protheroe \& Stanev (1993), and can generally be assumed
to be weak for plausible spectral indices $\alpha\ge 2$ above 30~TeV.
We ignore magnetic fields in the following calculations, although in a realistic scenario the
cascade emission will likely be isotropized due to the deflections
of the electrons and positrons by intergalactic magnetic fields.
Taking magnetic fields into account, we would have to extend
the luminosity function to include the unbeamed host sources of
blazars, the radio galaxies.  Their greater number density would
compensate for the lower luminosity resulting from the absence of
relativistic beaming.  We can thus approximate the cascade component
given by  

\begin{equation}
 \frac{dN^c}{dE_\gamma}(z_s) = \frac{(1+z_s)}{4\pi d_l^2(z_{s})}\
\int^{\gamma_{e,\rm max}}_{\gamma_{e,\rm min}}\left(\frac{dN_{\gamma_e, \epsilon}}{dt\ E_\gamma'}\right)
\left( \frac{dN_e}{d\gamma_e}\right) \ t_{IC}\ d\gamma_e
 \label{eq:cascade}
\end{equation}

\begin{equation}
 \frac{dN_{\gamma_e, \epsilon}}{dt\ E_\gamma'}(z_s)=\frac{2 \pi \ c\ r_0^2}{\gamma_e^{2}} \int_0^\infty  \frac{v(\epsilon, z_s)}{\epsilon}
f(x) d\epsilon
\label{eq:IC}
\end{equation}
with $f(x)=2x\log(x) + x + 1 - 2x^2
, \ \ (0<x<1)$
and $x=(E'_\gamma/4\gamma_e^2\epsilon)$.
The isotropic photon density of the cosmic microwave background
$v(\epsilon)$ is a black body with a temperature of
$T=(2.728\pm0.008)(1+z_s)$
\begin{equation}
v(\epsilon, z_s)=\frac{dN_{\epsilon}}{d\epsilon\ dV}=\frac{8\pi\epsilon^2}{h^3c^3}\frac{1}{\exp(\epsilon/kT)-1}.
\label{eq:cmb}
\end{equation}
In this calculation
positrons will be treated as electrons. On average, each absorbed gamma-ray
photon produces two electrons with an energy $\epsilon_e=1/2 E_\gamma^1$, since pair creation
has no preferred direction in the center-of-mass frame.
(Note that we use the index 1 to show that the gamma energy is different from the gamma energy in
Eq.~(\ref{eq:cascade}) and (\ref{eq:IC}).)
The electron spectrum
$dN_e/d\gamma_e$   as a function of the Lorenz factor $\gamma_e=\epsilon_e/mc^2$
for $\gamma_{\rm e, min}\ll \gamma_{\mathrm e} \ll \gamma_{\rm e, max}$, $\gamma_{\rm e, max}=30~\mathrm{TeV}/(2m_ec^2)$
and $\gamma_{\rm e, min}=$max$[\frac{1}{2}(E^1_\gamma/\epsilon_e)^{1/2}, 100~\mathrm{MeV}/(2m_ec^2)]$ is given by

\begin{equation}
\frac{dN_e}{d\gamma_e} = \frac{16\pi mc^2 d_l^2(z_{s})}{(1+z_s)}\ \frac{dN^i}{dE_\gamma^1}\ (1-e^{-\tau_{\gamma\gamma}(z_s)}).
\label{eq:elekspek}
\end{equation}

For numerical reasons the time integral has been replaced by an inverse-Compton time-scale $t_{IC}$ multiplication.
A comparison between the result using Eq.~(\ref{eq:cascade})-(\ref{eq:elekspek})  and a more exact calculation as in Protheroe \& Stanev (1993) shows
good agreement.

\begin{figure}

\vspace{0.7cm}

  \includegraphics[height=.25\textheight]{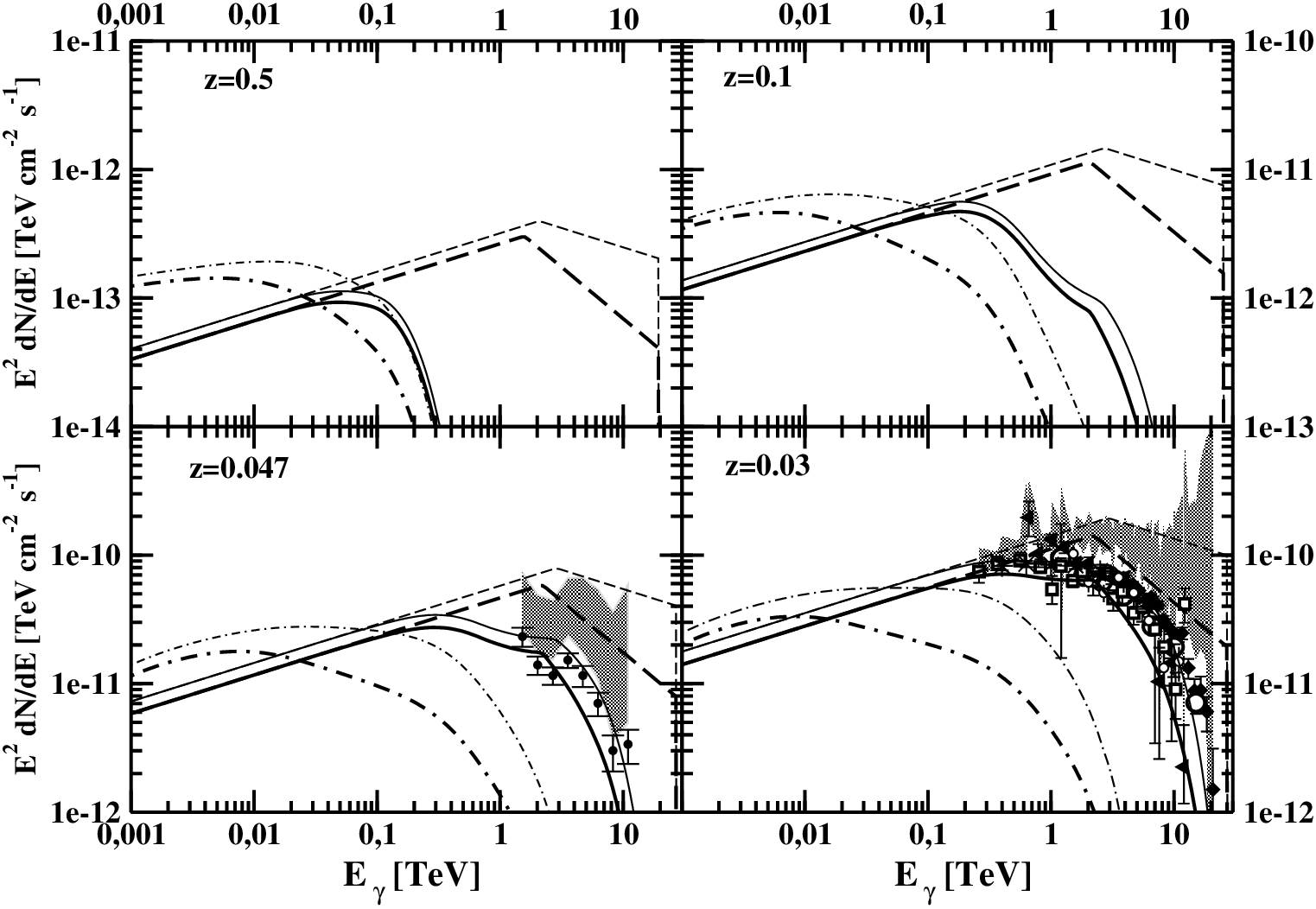}
  \caption{Template spectra for a source of case 1 (thin lines) and case 2
  (thick lines) located at different redshifts.
  For both cases the intrinsic spectrum (dashed lines), the
absorbed spectrum (solid lines) and the cascade emission (dot-dashed lines) is shown.
In the lower right panel the data are shown for Mkn501 (see Ref.
of Fig.1 and Kneiske et al. 2004). The shaded region is the intrinsic region
for different MRF models (Kneiske et al. 2004).  In the lower left panel the same
is shown for 1ES1959+650 at a redshift of 0.47.}
  \label{fig:lowMRF}
\end{figure}

\section{Undetected EGRET blazars}
The luminosity function of resolved EGRET sources, extended to the faint end,
has been computed by Chiang \& Mukherjee (1998).
We used their model changing only the spectral index from $\alpha=2.10$ to
$\alpha=2.37$. The new spectral index was determined by fitting the newly determined
EGRET background spectrum at energies $E_\gamma<1$~GeV. This choice is supported by averaging
the spectral indices for detected blazars from the 3rd EGRET catalog
(see Figure~\ref{fig:meanEGRET}) which led to a mean spectral index
of $2.37\pm0.04$.
We also included the cascade emission of the sources.
Although the total intensity of
the newly determined spectrum of the extragalactic background has become lower, the total contribution
of this blazar population (consisting mainly of flat-spectrum radio quasars)
to the extragalactic background in the energy range 100~MeV$<E_\gamma<$20~GeV
amounts to about 60\%.

\section{High peaked BL Lacs}
The remaining excess of the measured gamma-ray background
could be produced by the cascade emission of high-energy
peaked or X-ray blazars belonging to the HBL (XBL) class
(def. for HBL see e.g. Ghisellini et al 1998).
To test this idea we adopt an average HBL spectrum
and a TeV-luminosity function obtained from X-ray properties.

\subsection{Template spectra}

\begin{figure}

\vspace{0.7cm}

  \includegraphics[height=.24\textheight]{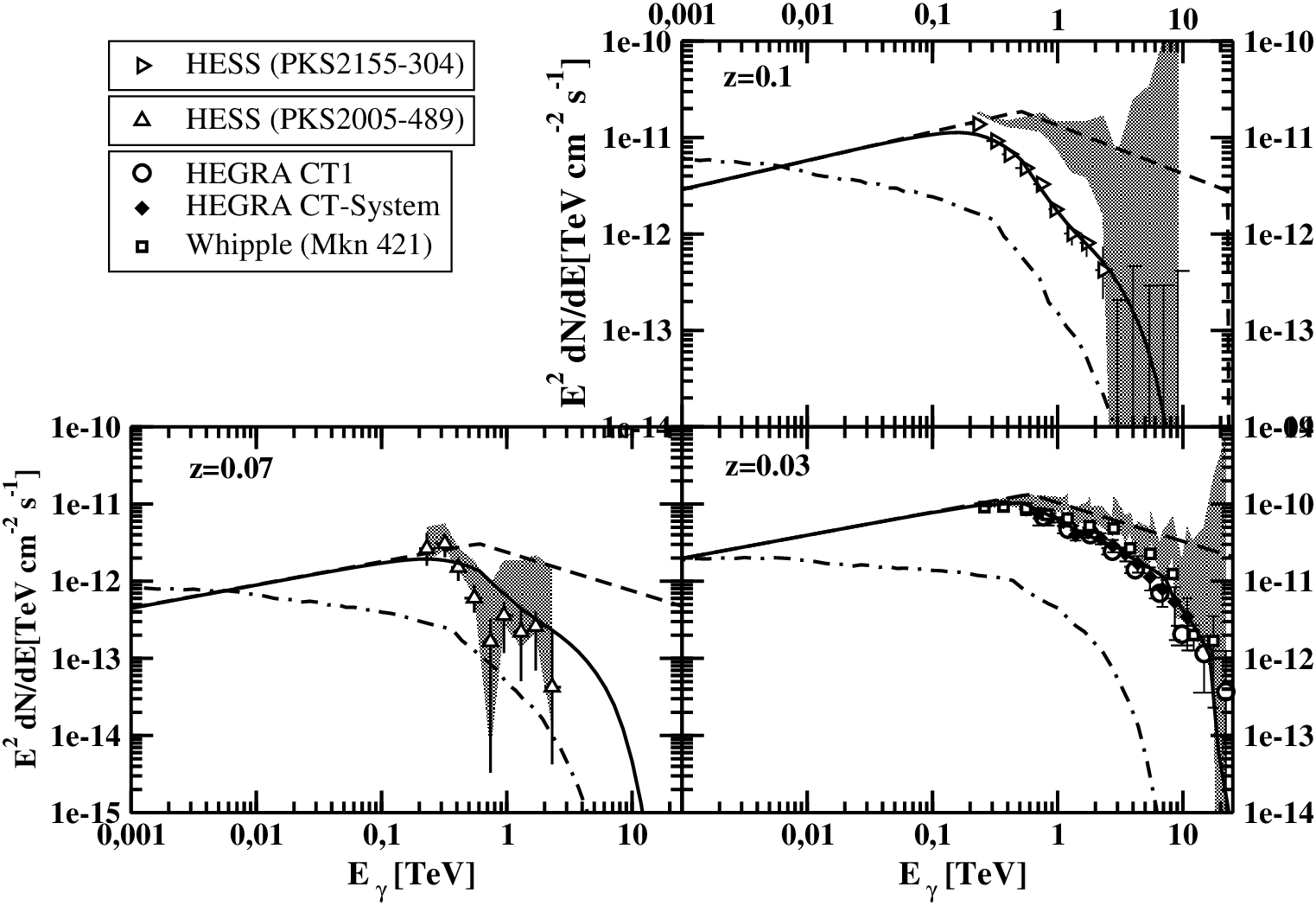}
  \caption{Same as in Figure~\ref{fig:lowMRF} but the flux is shown for
  a source of case 3. The data and intrinsic regions are for Mkn421, PKS2005-489 and PKS2155-304.
  (see refs Fig.1 and Kneiske et al. 2004).}
  \label{fig:intr}
\end{figure}

A number of extragalactic gamma-ray sources have been
detected with imaging air-Cherenkov telescopes (e.g., 
Wei 2006).
Some of them were bright enough to resolve their spectra in the TeV energy band.
At first glance the spectral energy distributions show a variety of shapes.
However, by calculating their intrinsic spectra taking the extragalactic absorption
into account, the spectra in the source frames are consistent with a remarkably similar
shape which can well be approximated with a broken power-law.
The spectral index above the peak is close to $\sim 2.5$ (cf. Mannheim 1998), while
it is much flatter below the peak ($<1.7$ in order to remain below the EGRET limits).

In order to assess the background contribution from cascading, it is important to constrain as
much as possible the template spectra and the metagalactic radiation field from observations.
Since the cascade emission depends on the maximum energy
and spectral index (Protheroe \& Stanev 1993), we try to adopt realistic templates for the spectra obtained
from observations.
The total flux of the cascade emission also depends on the photon density of the MRF
for which we will use models differing within the observational uncertainties.

We construct a template gamma-ray spectrum for Eq.~\ref{eq:Pow} by fitting a broken power-law
model to the data from detected BL Lacs.
For a set of $j$ data points with flux $dN^{obs}/dE_\gamma (z_{bl})$
of a source at redshift $z_{bl}$ and an optical depth for
gamma-rays in the Universe $\tau_{\gamma\gamma}(z_{bl})$ (10~GeV $< E_\gamma <$ 100~TeV)
the flux from the same source located at any given redshift $z=z_s$ is 
\begin{equation}
\left(\frac{dN^i}{dE_\gamma}\right)^j(z_s) = \frac{d_l^2(z_{bl})\ (1+z_s)}{d_l^2(z_s)\ (1+z_{bl})}
\ \left(\frac{dN^{obs}}{dE_j}\right)^j\ e^{\tau_{\gamma\gamma}(z_{bl})}.
 \label{eq:lumiflux}
\end{equation}

To test the influence of different, but realistic templates and the metagalactic
radiation field (MRF) we will discuss four different cases:
\begin{enumerate}
\item A template obtained from observations of Mkn501 and 1ES1959+650 and a high MRF model.
These are standard sources with similar slopes around the peak energy.
\item A template obtained from  observations of Mkn501 and 1ES1959+650 and a low MRF model, to test
the uncertainty of the MRF.
\item A template obtained from  observations of Mkn421, PKS2005-489, PKS2155-304  and a high MRF model.
These sources show a steeper decline at higher energies, which could be a sign of intrinsic absorption.
\item A template obtained from  observations of H1426+428 and a high MRF model, to show the effect of an
extremely high value for the peak energy.
\end{enumerate}
\begin{table}
\centering
\caption{Parameters of HBL template spectra.}
 \begin{tabular}{llll}
 \hline \hline\
 model & $E_p$ [TeV] & $\alpha_1$ & $\alpha_2$\\
 \hline
 1 & 3.0 & 1.7 & 2.3 \\
 2 & 2.3 & 1.7 & 2.8 \\
 3 & 0.6 & 1.7 & 2.5 \\
 4 & 10.0 & 1.2 & 2.0 \\

 \hline
 \end{tabular}
 \label{tab:HBLpara}
 \end{table}
The parameters of the HBL templates are the two spectral
indices and the location of the maximum (the normalization is obtained from
the luminosity function, see below).
The numbers in Table~\ref{tab:HBLpara} are the result of fits of the
intrinsic spectral energy distribution that have been calculated using Eq.~(\ref{eq:lumiflux})
and the MRF model presented in Kneiske et al. (2002, 2004).

For the high MRF we have assumed the
"high-UV" model including the "warm-dust" component. The low MRF is the
"low-IR" model without any UV emission.
The cascade emission from an average HBL is then given by
Eq.~(\ref{eq:cascade})-(\ref{eq:elekspek}).
The results can be seen in Fig.~\ref{fig:lowMRF} and Fig.~\ref{fig:intr}.

In Fig.~\ref{fig:lowMRF} the intrinsic spectra (dashed lines), the
absorbed spectra (solid lines) and the cascade emission (dot-dashed lines) are
plotted for a source of case 1 (thin lines) and case 2 (thick lines).
The four different panels show the change in the average spectrum
if the HBL is located at different redshifts. Due to the
increasing distance the flux is smaller and the
extragalactic absorption higher.
For comparison the data and allowed regions for the intrinsic spectra of Mkn501
are shown in the lower right panel at a redshift $z=0.03$ and for 1ES1995+650
in the lower left panel at a redshift $z=0.047$.

In Fig.~\ref{fig:intr} the same is plotted as in Fig.~\ref{fig:lowMRF}
except for a source of case 3. The data and allowed regions for the intrinsic spectra are taken
from Mkn421, PKS2005-489 and PKS2155-304 and plotted in the panel depending on their
redshift range. Within the uncertainties of the data all sources
in the same figure can be fitted roughly with the same template.

\begin{figure}

\vspace{0.7cm}

  \includegraphics[height=.24\textheight]{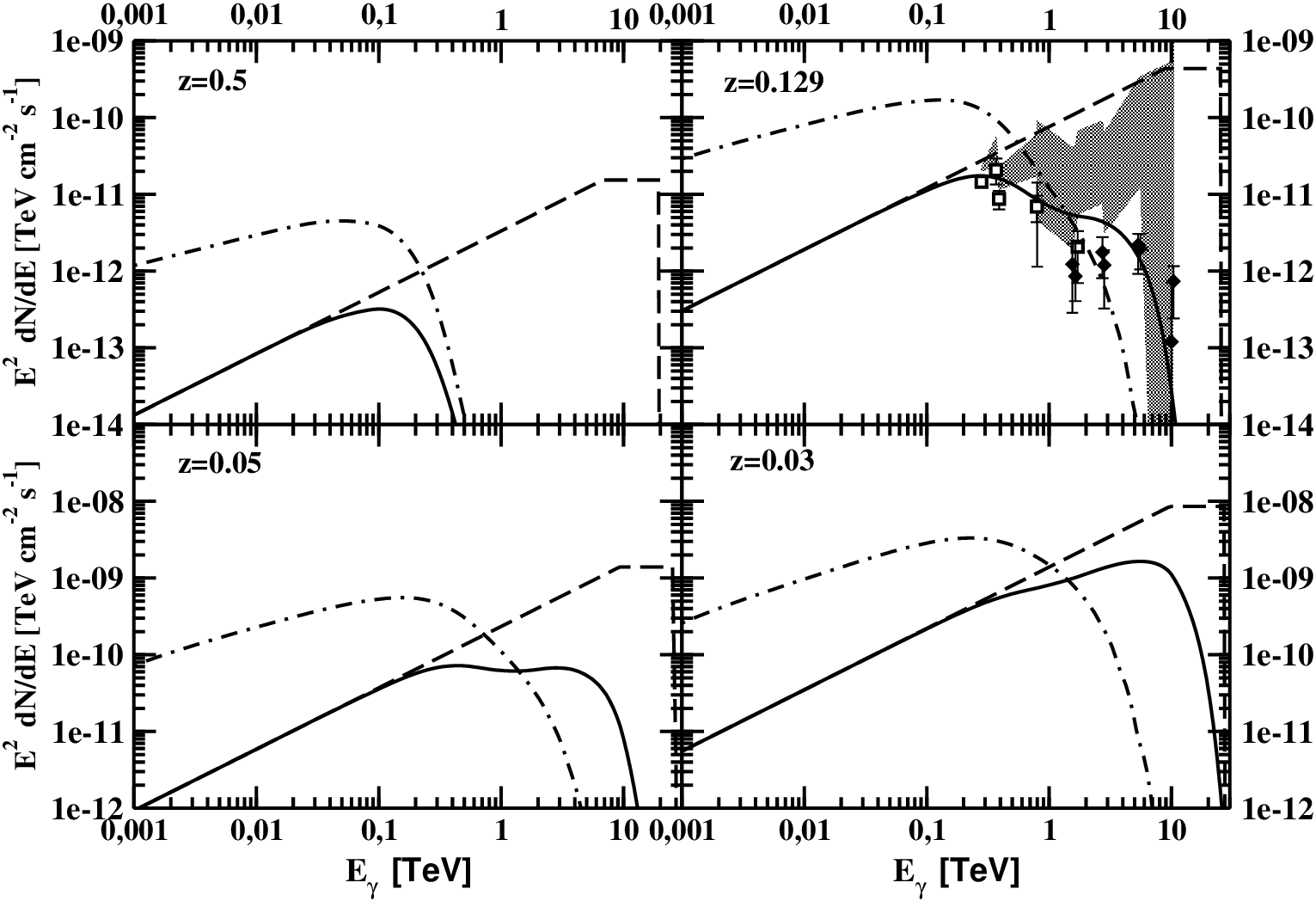}
  \caption{The un-absorbed spectrum (dashed line), the
absorbed spectrum (thick solid line) and the cascade emission (dot-dashed line)
for the extreme BL Lac 1H1428+428  at a redshift of $z=0.129$.
The data and the shaded region indicating the allowed range for the intrinsic spectrum is taken from Kneiske et al. (2004). }
  \label{fig:ExBL}
\end{figure}

The location of the peak energy in the synchrotron spectrum of blazars apparently varies across a wide range from
the near-infrared (LBLs) to X-rays (HBLs), although thermal emission from heated dust and starlight
as well as photoelectric absorption in the EUV and soft X-rays hampers an unbiased measurement of the non-thermal spectrum.
Blazars such as H1426+428 seem to exhibit, at least occasionally, a peak energy beyond $\sim 100$~keV, and a gamma-ray peak
beyond $\sim 10$~TeV  when accounting for extragalactic absorption (Costamante et al. 2001 and 2003, Kneiske et al. 2004).
In Fig.~\ref{fig:ExBL} we show the effects of cascading on such a spectral energy distribution (SED),
adopting $E_p$=10~TeV,  $\alpha_1=0.2$ and $\alpha_2=1.0$.
The duty cycle of the extreme behavior is unknown, and we consider a sizeable contribution of such extreme BL Lacs as rather speculative.
In Fig.~\ref{fig:ExBL} the same is plotted as in Fig.~\ref{fig:lowMRF}
except for a source of case 4 and data of H1426+428 at a redshift of $z=0.129$.

\subsection{TeV-luminosity function}

The luminosity function at gamma-ray energies (TeV-LF) of HBL is poorly known, since there has not been
a complete survey, and the number of known sources is still rather low.
We construct a LF based on the somewhat better known X-ray properties of
HBLs.
Bade et al. (1998) and Laurent-M\"uhleisen et al. (1999) obtained
a luminosity function based on the samples from the ROSAT All Sky Survey,
Rector et al. (2000) and Caccianiga et al. (2002) used samples from the Einstein Medium Sensitivity Survey
and the Radio Emitting X-ray Sources catalog.
For this study, we used the LF of
Beckmann et al. (2003) who combined all the available data and derived a LF for BL Lacs.
To obtain the TeV-LF we assume the luminosity between 0.5~keV to 2~keV $L_{(X)}$
equals the gamma-ray luminosity above 0.3~TeV  $L_{(\rm TeV)}$. This assumption is
based on the statistics from 246 sources using an SSC model and relations from blazar observations
(Costamante \& Ghisellini 2002).
A minimum, maximum and break luminosity of $L_{(\rm min)} = 10^{43}$ erg s$^{-1}$,
$L_{(\rm max)} = 10^{47}$ erg s$^{-1}$ and $L_{(\rm br)} = 10^{45}$ erg s$^{-1}$ respectively
is used.

\section{Results}

\begin{figure}

\vspace{0.7cm}

  \includegraphics[height=.24\textheight]{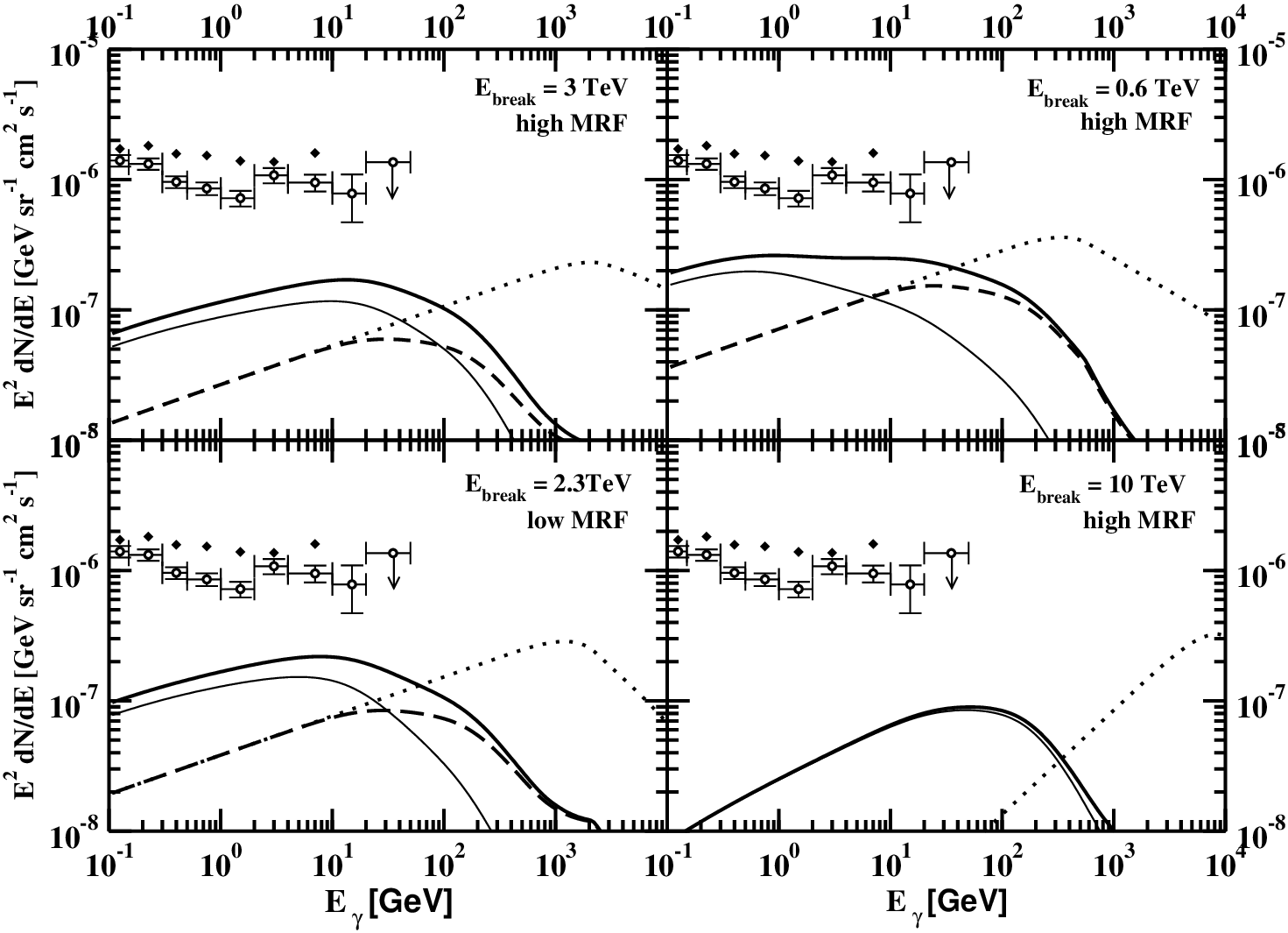}
  \caption{The intensity of the extragalactic gamma-ray background.
  The contribution from HBLs to the EGRB (thick solid line) for all 4 cases.
  (The flux without absorption: dotted line; flux with absorption: dashed line;
  cascade contribution alone: thin solid line). Data are from Sreekumar et al. (1998): solid diamonds
  and Strong, Moskalenko \& Reimer (2004): open circles.}
  \label{fig:gammaback}
\end{figure}

The contribution of the HBL component for all 4 cases is shown in
Fig.~\ref{fig:gammaback}.
Without the effect of absorption
and the resulting cascade emission the contribution to the background would
be small. The cascade emission enhances the background intensity
by up to an order of magnitude in the EGRET energy range.
HBL can thus contribute 7\%, 11\%, 19\% or 1\% to the gamma-ray background  in the
energy range 100~MeV$<E_\gamma<$20~GeV for case 1 to case 4 respectively.

\begin{figure}

\vspace{0.7cm}

  \includegraphics[height=.24\textheight]{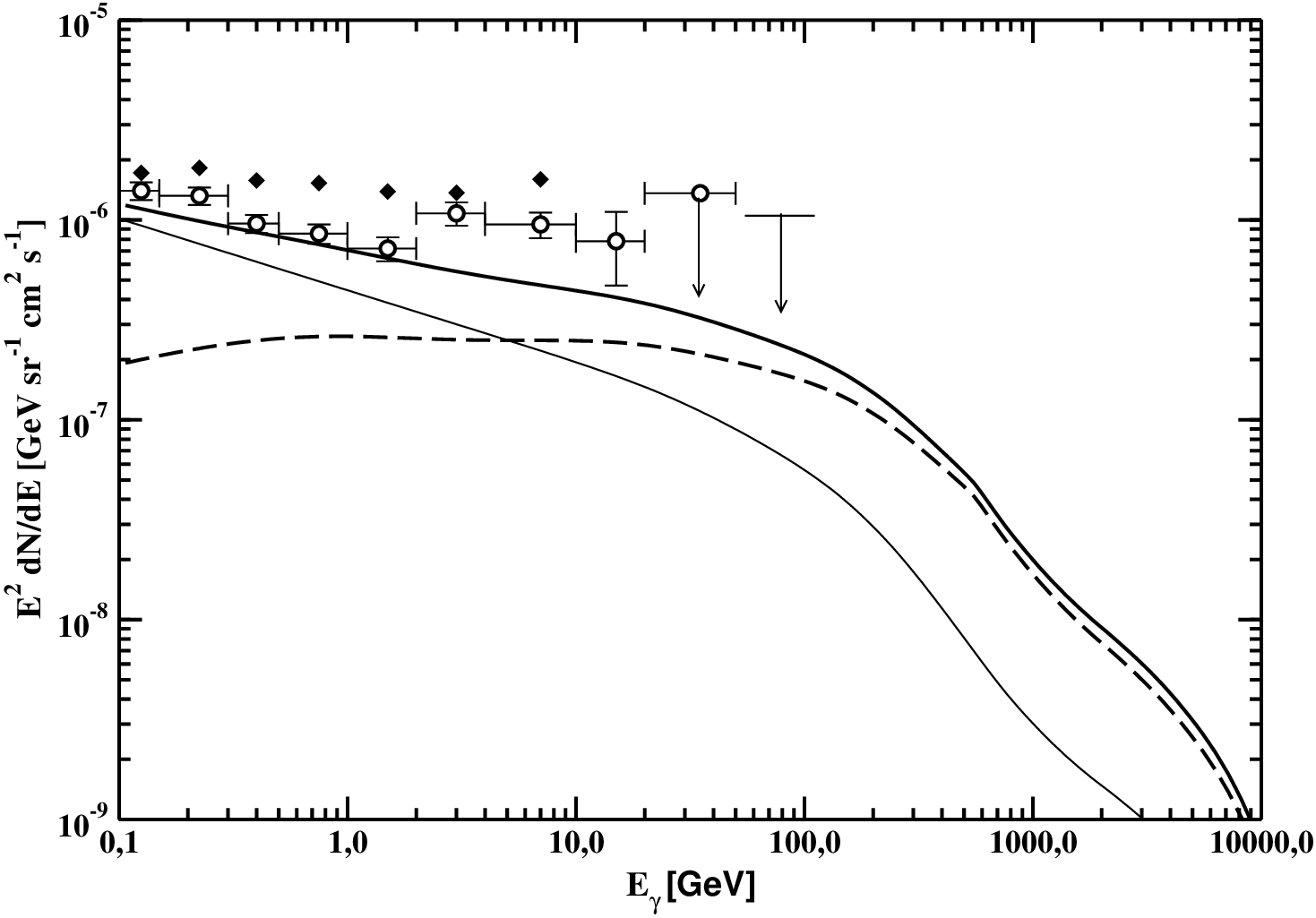}

  \vspace{0.5cm}

  \caption{Extragalactic gamma-ray background and assumed blazar contributions due to
  low-peaked blazars (thin solid line) and high-peaked blazars (dashed line)
  including the effects of intergalactic cascading.}
  \label{fig:gammaback2}
\end{figure}

For comparison, the contribution from EGRET blazars (LBL, FSRQ) is shown together
with the result for case 3 in Fig.~\ref{fig:gammaback2}.
Due to the new spectral index, the change of cosmological parameters,
the additional cascade emission and the reanalyzed EGRET data, the unresolved
EGRET blazars now produce $\sim60$\% of the background intensity.
Comparing the total intensity as a sum of the two contributions (thick solid line)
and the EGRET data, the agreement is acceptable at energies below 2~GeV but is too
small above this energy.

Using
\begin{equation}
\frac{dN}{dz}(z)=\int^{\infty}_{L_{\gamma, {\rm min}}}
\ \frac{dV}{dz}\frac{dN}{dV dL_{\gamma}}dL_{\gamma}
\label{eq:dNdz}
\end{equation}

and

\begin{equation}
L_{\gamma, {\rm min}}(z) = \frac{4\pi d^2_l(z)\ F_{lim,\gamma} }{(1+z)}
\end{equation}

\begin{figure}

\vspace{0.7cm}

  \includegraphics[height=.24\textheight]{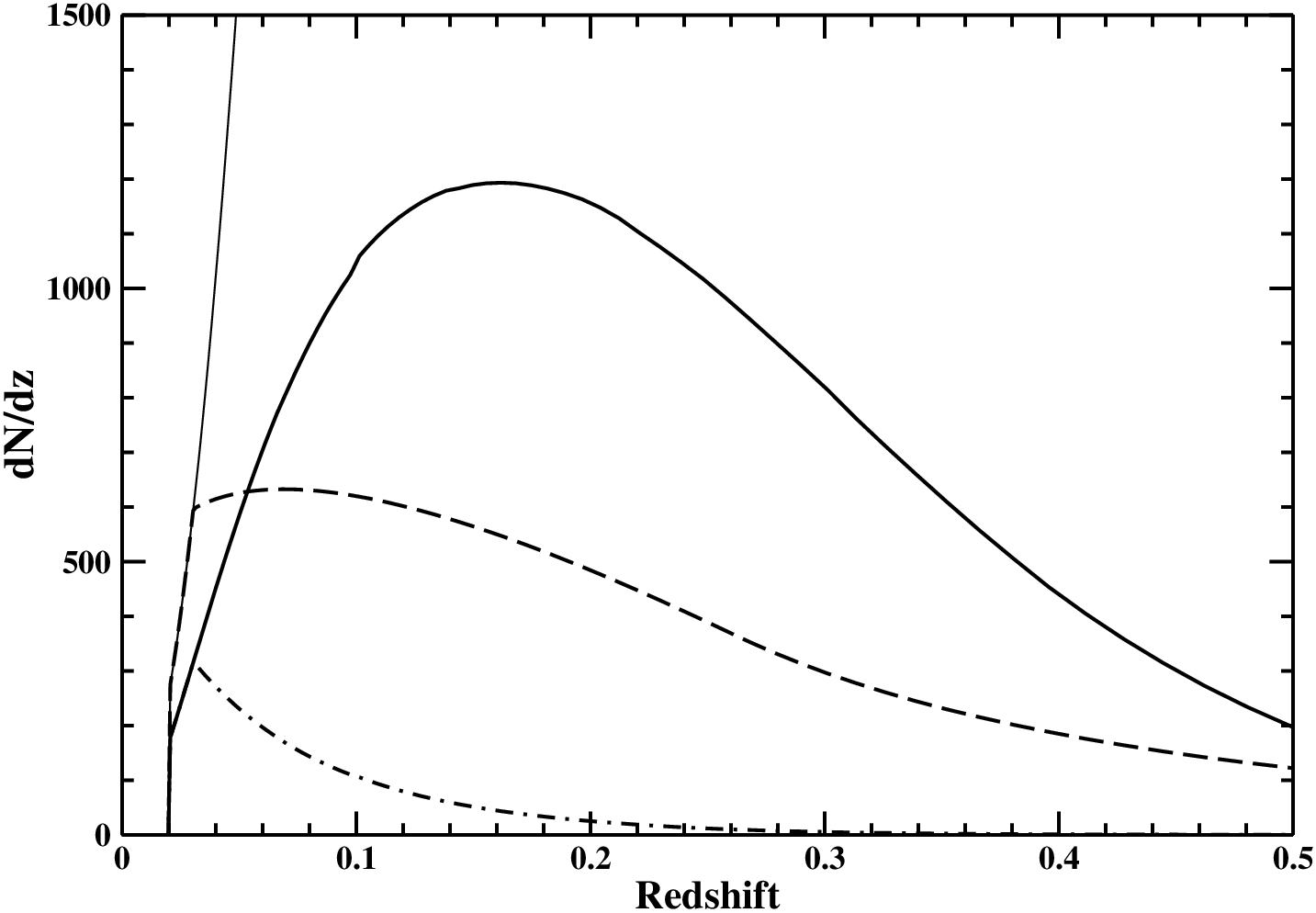}

\vspace{0.5cm}

  \caption{Number of BL Lacs as a function of redshift for case 1; thin
  solid line: total number of sources; thin dashed line: observable sources using a telescope $E_{thr}=300$~GeV and a flux limit of
  $10^{-11}$cm$^{-2}$ s$^{-1}$; thick-solid line: observable sources including
  the effect of extragalactic absorption only; dot-dashed line: observable sources including the
  effect of extragalactic absorption and telescope parameters (energy threshold and sensitivity).}
  \label{fig:dNdz}
\end{figure}

we can check the number counts of HBLs above 300~GeV which could be observed above a flux limit
of $10^{-11}$ cm$^{-2}$ s$^{-1}$ according to our choice of the LF
with and without extragalactic absorption (see dot-dashed and dashed line in Fig.~\ref{fig:dNdz}, for case 1).
The total population consists of about $10^6$ HBLs for $0.02<z<5$ (thin-solid line).
We obtain 23, 24, 30 or 22 sources at redshifts $z<0.3$ for case 1 to case 4 respectively.
The peak redshift of the observed sources would be $z = 0.03$.
The results are slightly higher than the number counts obtained from HBL observations with HEGRA and Whipple.
However we did not include the change of flux limit due to different zenith angles or other
observational based effects.
The number of detectable sources with the
new Cherenkov telescopes (MAGIC, H.E.S.S., Veritas, Cangaroo)
should increase by a factor of 10 observing at lower energy and in both hemispheres, which is in agreement
with the $\approx$ 100 HBL from integration of the X-ray luminosity function within one Gpc.
From Fig.7 it is clear that independent of the telescope parameters the maximum number of sources
will be detected at a redshift of $\approx 0.16$ (thick solid line).

\section{Discussion}

The EGRET extragalactic gamma-ray background data may be affected by large
systematic errors. They are strongly dependent on the galactic foreground emission, which
must be subtracted from the measured signal.
Improvements in the modeling of the foreground emission
were the reason for the new determination of the extragalactic background by Strong, Moskalenko \& Reimer (2004).
Recently, large amounts of gas, possibly connected
with the Gould Belt, which are not accounted for in existing HI and CO
surveys, have been found using infrared and gamma-ray observations.
This implies severe revisions of the gamma-ray interstellar emission models to high latitudes.
Grenier, Casandijan \&Terrier (2005) obtained a new estimate of the
extragalactic gamma-ray background including this newly found interstellar matter.

The largest uncertainties in determining the blazar contribution to the extragalactic background
are their unknown SED and LF where only coarse estimates could be derived. In particular, at the
low luminosity end of the LF, blazars could hide inside elliptical galaxies.
At the present stage, the shape of the intrinsic spectra of blazars, as well as the number of
blazars, seem to suffice to produce most of the background from faint, unresolved sources.
This has also been discussed in Giommi et al. (2006). They have used 5 GHz radio number counts
and a SSC model to calculate
the contribution of blazars to the CMB, X-ray and gamma-ray background. They extrapolate
the number counts to very low fluxes below 1 Jansky. As a result they found that
the high energy gamma-ray background is overproduced by blazars if their SED shows an all-time
high-energy gamma-ray bump. It is not clear if all
faint radio sources which are included in their sample are producing gamma-rays and including
these sources could lead to an overestimate of the background flux at gamma-ray energies.
A more accurate treatment could start from generalizing the known correlations between
X-ray/gamma-ray peak and luminosity for a single, luminosity-dependent template using leptonic
and hadronic models.
It is not clear whether accelerated electrons or hadrons (or both) are responsible
for the gamma-ray emission. Both models can reproduce the multi-wavelength data equally well, since
not enough simultaneous data are available (Aharonian et al. 2005).
Therefore we have used simple power-laws for the spectral energy distribution.
Recent observations at GeV and TeV energies have also shown flares in blazars which are
only bright in gamma-rays. These so-called orphan flares have to be explained by more complex models
(Reimer, B\"ottcher \& Postnikov 2005).
To obtain more precise blazar SED templates, detailed modeling of single sources with leptonic and hadronic models
are needed, which will be possible if more simultaneous data from multi-wavelength campaigns are available.
Number counts of HBLs found with IACTs will also be crucial to test the assumed LF and templates.

Blazars show variability across a wide band of time scales, from minutes to years, and estimates
of their contribution to the extragalactic background must take into account selection
effects due to the excess of flaring sources among detected sources found in incomplete
surveys, and on the shape of the SED.  Since the duty cycle of flares seems to be rather
small, we have ignored such selection effects in this work.

Another source of uncertainty in the above calculations is the strength of the
 magnetic fields in the blazar host galaxies, in clusters of galaxies, and
in intergalactic space. The
magnetic field strength in the Coma cluster of galaxies has been measured
to be (1.7$\pm$0.9)~$\mu$G (see Kronberg 1994
for a review). The cluster magnetic field seems to be correlated also with larger
filaments of the matter concentration in the Universe.
Using rotation measurements of radio sources Ryu, Kang \& Biermann (1998)
could show that the magnetic fields in these regions could be as strong as  $\simeq 1~\mu$G.
For the gamma-ray templates considered in this work,
the mean free path takes values between a few hundred Megaparsec and some Gigaparsecs, enough to
assure that the gamma-rays escape freely from the intracluster medium before pair production
occurs.  In general, they would then enter void regions, unless the jet directions are
correlated with the supergalactic filament orientation.
The magnetic field in void regions is not well known, but it seems that it drops to $10^{-9}$~G
or less (Lee, Olinto\& Sigl 1995).
Detailed cascade simulations and observations of gamma-ray halos can help to set
limits on the extragalactic magnetic field. Aharonian et al. (2001) found no sign of a halo
in Mkn501 in its quiescent phase. This would lead to much smaller magnetic fields like $10^{-16}$~G.
The calculations presented above were made assuming no magnetic field. The effects of weak magnetic fields ($10^{-17}-10^{-20}$G)
on cascade emission in Mkn501 and H1426+428 are discussed in Dai et al. (2002) and Fan, Dai \& Wei (2004),
respectively. For Mkn501 magnetic fields up to $10^{-18}$~G have almost no influence on the cascade emission at
energies above 100~MeV. A magnetic field of $10^{-17}$~G reduces the flux by almost one order of magnitude
at 100~MeV. Comparing this result with the calculations of H1426+428 the effect seems to depend strongly on 
the intrinsic spectrum and redshift of the BL Lac. Therefore an estimate 
for the gamma-ray background seems
only possible by an exact simulation including the extragalactic magnetic field as a parameter.
This will be done in future work.

Stronger magnetic fields ($>10^{-12}$~G) would lower the cascade flux 
from the blazars emitting beamed radiation by about a factor of hundred. 
However, the correspondingly larger number of blazar host galaxies
(radio galaxies) would then come into play to compensate for the loss in flux per source.

An analysis by Gorbunov et al. (2005) suggested a very weak correlation between
the diffuse photons above 10 GeV detected by EGRET and some BL Lacs (LBL and HBL).
The low number of blazars to which the photons can be traced back might indicate
 a greater value of the intergalactic magnetic field strength, sufficient to isotropize
the cascade emission.

The hard feature in the background spectrum at a few GeV
might also indicate other emission components, with less broad continuum radiation.
It has been suggested that neutralino annihilation in clumpy dark matter halos
might naturally give rise to this bump (Els\"asser and Mannheim 2005).

Stawarz, Kneiske \& Kataoka (2005) studied another possible contribution
from the parsec scale jets of
Faranoff-Riley type I galaxies. The observed X-ray emission in the knots of the extended jets
could result in an inverse Compton component at gamma-ray energies.
But neither the direct emission nor the faint cascade emission will lead to a larger
contribution than 1\%. Using a consistent model where the TeV emission is produced
in the inner core region and in the extended jets as well could increase the
contribution of FR-I galaxies to the extragalactic gamma-ray background
(for the case of M87, see Reimer, Protheroe \& Donea 2004, Stawarz et al. 2005).

Based on the observational finding of an increasing number of HBLs detected above 100~GeV energies,
and the inescapable effect of intergalactic cascading on their emission, we have shown
that they can indirectly contribute a sizeable fraction of the extragalactic gamma-ray background below 100~GeV.
Ongoing observations with IACTs (e.g., H.E.S.S., MAGIC, VERITAS)
will allow one to probe the assumed source population at very high energies.
Monitoring campaigns will be important to obtain the time-averaged flux emitted by the
blazar population at gamma-ray energies, and to better understand the selection
effects due to flaring sources. A significant increase
in source statistics will be needed to test the assumed luminosity function.

\section{Acknowledgments}
This research was supported by the BMB+f under grant 05AM9MGA.
We thank Ray Protheroe for helpful discussions and the anonymous referees for
comments and suggestions.

\end{document}